\documentclass[a4paper,fleqn,usenatbib]{mnras}

\usepackage{newtxtext,newtxmath}

\usepackage[T1]{fontenc}
\usepackage{ae,aecompl}

\usepackage{graphicx}	
\usepackage{amsmath, bbold}	
\usepackage{amssymb}	
\usepackage{dsfont}		
\usepackage{trfsigns}	
\title[On the dynamics of Comet 1P/Halley]%
{On the dynamics of Comet 1P/Halley: Lyapunov and power spectra}

\author[Jorge A. P\'erez-Hern\'andez and Luis Benet]{
Jorge A. P\'erez-Hern\'andez$^{1}$\thanks{E-mail: jperez@icf.unam.mx}
and Luis Benet$^{1}$\thanks{E-mail: benet@icf.unam.mx}
\\
$^{1}$Instituto de Ciencias F\'isicas,
Universidad Nacional Aut\'onoma de M\'exico, UNAM,
Apdo. Postal 48-3, 62251 Cuernavaca, Mor., Mexico\\
}

\date{Accepted 2019 April 16. Received 2019 March 26; in original form 2018 December 7.}

\pubyear{2019}

\begin{document}
\label{firstpage}
\pagerange{\pageref{firstpage}--\pageref{lastpage}}
\maketitle

\begin{abstract}
Using a purely Newtonian model for the Solar System, we investigate
the dynamics of comet 1P/Halley considering in particular the Lyapunov
and power spectra of its orbit, using the nominal initial conditions
of JPL's Horizons system.
We carry out precise numerical integrations of the $(N+1)$-restricted
problem and the first variational equations, considering a time
span of $2\times10^5$~yr.
The power spectra are dominated by a broadband component, with
peaks located at the current planetary frequencies, including contributions
from Jupiter, Venus, the Earth and Saturn, as well as the $1:6$
resonance among Halley and Jupiter and higher harmonics.
From the average value of the maximum Lyapunov exponent we estimate the
Lyapunov time of the comet's nominal orbit,
obtaining $\tau_L \simeq 562$~yr; the remaining independent
Lyapunov exponents (not related by time-reversal symmetry)
tend asymptotically to zero as $t^{-1/2}$.
Yet, our results do not display convergence of the maximum Lyapunov
exponent. We argue that the lack of convergence of the maximum Lyapunov
exponent is a signature of transient chaos which will lead to an eventual
ejection of the comet from the Solar System.
\end{abstract}




\section{Introduction}

Comet 1P/Halley (or comet Halley) is one of the oldest observed comets of the
Solar System, with observations already recorded back to
240~BC~\citep{1986JRASC80_62Y}. The (approximate) orbital period of the comet
is about $75.3$ years~\citep{giorgini1996jpl}. The Lyapunov time (inverse of the maximal
Lyapunov exponent) has been the subject of attention of recent works.
Based on previous work by~\citet{chirikov1989chaotic},
which models the dynamics of comet Halley at every pericenter passage by a 2-d
discrete map, \citet{shevchenko_2006} obtains the Lyapunov time
$\tau_L \sim 34$~yr. More recent calculations
using $N$-body numerical integrations of Newtonian models for
the Solar System \citet{munoz2015chaotic} obtain $\tau_L \sim 70$~yr, while
\citet{boekholt2016origin} obtain $\tau_L \sim 300$~yr. From now on we shall
concentrate on the last two time scales, simply because they are obtained
from $N$-body integrations, despite of differences
in the models, initial conditions and the numerical methods employed.

The Lyapunov time $\tau_L$ defines the so-called $\e$-folding time,
a time scale for which an initial small deviation grows by a
factor of $\e$. To illustrate the significance of the
Lyapunov times quoted above,
consider an initial small deviation $\delta_0 = 150$~km, which is of the order
of magnitude of the uncertainty in the semi-major axis of comet Halley during
the last perihelion passage~\citep{1986A&A...163..246L}. Then, the time required
for that initial deviation to grow to a distance comparable with Earth's radius
is $262$ and $1123$ years, respectively, using the two time scales quoted above.
These time scales seem to be rather short to account for the time span of the
recorded observations of comet Halley, which have been
succesfully reproduced~\citep{yeomans1981long}.

Aside from the differences in the Solar System models and the numerical
integration schemes, both \citet{munoz2015chaotic} and \citet{boekholt2016origin}
use fiducial trajectories to estimate the maximum Lyapunov
exponent~\footnote{We note that throughout this paper we shall refer
to the Lyapunov exponents, though in practice our calculations and
statements refer to the {\it finite-time} Lyapunov numbers.} of
Halley's orbit, again, with some differences in the implementation.
\citet{2001AJ....121.1171T} pointed
out the possibility to overestimate the maximum Lyapunov exponent when using
``two particle'' methods, and that those problems are absent when using
variational methods to calculate the exponent. This motivates the present study.

In this paper, we present results on the computation of the full Lyapunov
spectrum of comet Halley integrating numerically the equations of motion
together with the (first) variational
equations~\citep{benettin1980lyapunovI,benettin1980lyapunovII}.
Our results, using a purely Newtonian model, the initial conditions of JPL's Horizons
system corresponding to February 17th, 1994, 00:00:00.0 (TDB) and
an integration spanning $2\times10^5$~yr, suggest an inverse average
maximal Lyapunov exponent of $\sim 562$~yr. This time scale is larger than
those reported earlier. We also find
that the remaining independent Lyapunov exponents tend to zero asymptotically,
seemingly as $t^{-1/2}$. Yet, our results are not conclusive with regards to
attaining convergence of the maximum Lyapunov exponent which, we argue,
is related to an eventual escape of the comet from the Solar System.

Our paper is organized as follows: In Section~\ref{Sec2} we describe the specific
Newtonian model used to study the dynamics of comet Halley, which is akin to a
restricted spatial $(N+1)$-body problem, and the associated equations of motion.
Section~\ref{Sec3} is devoted to the description of the numerical methods used
to integrate the equations of motion as well as the variational equations to
obtain the Lyapunov spectrum. We use the conservation of the total energy of
and the conservation of the $z$-component of the angular momentum of the
$N$ Solar System bodies of our model to test the quality of our integration;
we also use the sum of all Lyapunov exponents to show that our integration
behaves numerically as being symplectic. In Section~\ref{Sec4} we describe our
numerical results, addressing in particular the semi-major axis, eccentricity
and inclination of Halley's orbit, the power spectra related to these quantities,
and Lyapunov spectrum. Finally, in Section~\ref{Sec5} we discuss how our results
compare to those of \citet{munoz2015chaotic} and \citet{boekholt2016origin}, and
in Section~\ref{Sec6} we present our conclusions.

\section{Dynamical model for Comet Halley}
\label{Sec2}

We consider a purely Newtonian model for the dynamics of the Solar System. More
specifically, we consider the $N$-body Newtonian dynamics of the Sun, all planets,
and the Moon, all modeled as point-particles. Comet Halley is modeled as a
massless test-particle interacting gravitationally with $N=10$ members of the
Solar System, without affecting them. Therefore, we are considering a spatial
restricted $10+1$-body problem. Since there is evidence of past
close-approaches of comet Halley to the Earth-Moon
system~\citep{yeomans1981long}, we consider these bodies separately in our
model.

The equations of motion associated to the $N$-bodies of the Solar System
($j = 1, \ldots,10$, counting from the Sun outward with respect to the
nominal semi-major axis and the Moon being placed after the Earth) can be
written as
\begin{eqnarray}
\label{Eq:vel}
\dot{\mathbf{r}}_j   & = & \mathbf{v}_{j},\\
\label{Eq:acc}
\dot{\mathbf{v}}_{j} & = & - \sum_{i=1,i\neq j}^{N}
\frac{G m_i \mathbf{r}_{i,j}}{r_{i,j}^3}.
\end{eqnarray}
Here, $\mathbf{r}_j$ denotes the position vector in cartesian coordinates
of the $j$-th body, $\mathbf{v}_j$ denotes its velocity,
$\mathbf{r}_{i,j}=\mathbf{r}_i-\mathbf{r}_j$ is the relative position
vector of particle $j$ with respect to the $i$-th particle,
$r_{i,j} = |\mathbf{r}_{i,j}|$ is their mutual distance, and $G$ is
the gravitational constant. The origin of coordinates
corresponds to the barycentric center of mass of the Solar System.
As it is clear from Eqs.~(\ref{Eq:vel}) and~(\ref{Eq:acc}), any relativistic
corrections to the orbit of Mercury, and also to the orbit of comet Halley,
are neglected.

These equations of motion conserve the total energy
\begin{equation}
\label{Eq:energy}
E = \frac{1}{2}\sum_{i=1}^N m_i |\mathbf{v}_i|^2
- \sum_{i,j,i\neq j}^{N} \frac{G m_i m_j}{r_{i,j}},
\end{equation}
and the total angular momentum vector
\begin{equation}
\label{Eq:angular-momentum}
\mathbf{L} = \sum_{i=1}^N m_i \mathbf{r}_i \times \mathbf{v}_i.
\end{equation}
The conservation of these quantities shall be used below to illustrate the
accuracy of our numerical integrations.

Considering comet Halley as a massless point particle, its equations of motion
have the same structure as Eqs.~(\ref{Eq:vel}) and~(\ref{Eq:acc}). In this case,
we write
\begin{eqnarray}
\label{Eq:Hvel}
\dot{\mathbf{r}}_H   & = & \mathbf{v}_H,\\
\label{Eq:Hacc}
\dot{\mathbf{v}}_H  & = & - \sum_{i=1}^{N}
\frac{G m_i \mathbf{r}_{i,H}}{r_{i,H}^3}.
\end{eqnarray}
We notice the explicit appearance of time through the positions of the $N$
Solar System bodies.

\section{Numerical methods}
\label{Sec3}

The differential equations described above, Eqs.
(\ref{Eq:vel})-(\ref{Eq:acc}) for the Solar System and (\ref{Eq:Hvel})-(\ref{Eq:Hacc})
for comet Halley, are integrated
simultaneously using Taylor's method, as implemented in \texttt{TaylorIntegration.jl}
\citep{taylorintegration}, which also permits to compute the full
Lyapunov spectrum. The idea of Taylor's method is to approximate locally
the solution of the equations of motion by means of
high-order Taylor expansions of the dependent variables in terms of the time $t$.
The coefficients
are computed recursively in each time step, exploiting automatic differentiation
techniques. Taylor's method allows for integrations with high accuracy,
achieving essentially round-off errors per integration
step~\citep{jorba2005software}. This is obtained by using a high-enough order of the
polynomial expansion, which is computationally more efficient than
using smaller time steps. In the results presented below, we use
polynomials of order $28$, and choose the time step by imposing
that the last two terms of the expansion are in absolute value smaller
than $\varepsilon=10^{-20}$. Since the time steps are not constant, the method
has an adaptative step size.

In order to compute the full Lyapunov spectrum of Halley, we
follow~\citet{benettin1980lyapunovII}, which is based on Oseledet's
multiplicative ergodic
theorem~\citep{Oseledet1968multiplicative,benettin1980lyapunovI}; see
also~\citet{skokos2010lyapunov} or \citet{pikovsky2016lyapunov} for recent treatments of the
subject. Together with the integration of the equations of motion, we integrate
the first-order variational equations associated to variations of Halley's
coordinates and velocities. More concretely, considering
the time $t$ as the independent variable and $x \in \mathbb{R}^{d}$ as the dependent
variables of the equations of motion, which we write as $\dot{x} = f(t,x)$, the
first-order variational equations are given by
\begin{equation}
    \dot \xi = (\operatorname{D}f)(x(t))\cdot \xi,
	\label{eq:general_var_eqs}
\end{equation}
where $(\operatorname{D}f)(x(t))$ is the Jacobian of $f(t,x)$ with respect
to the dependent variables $x$, evaluated at time $t$, obtained around a
given solution
$x(t)$ of the equations of motion. Here, $\xi$ represents the deviations from the
nominal solution $x(t)$ due to infinitesimal changes in the initial conditions
of Halley's orbit.

From the equations of motion of Halley's comet,
Eqs.~(\ref{Eq:Hvel}) and~(\ref{Eq:Hacc}),
the Jacobian $\operatorname{D}f$ associated to Halley's phase space variables
is given by
\begin{equation}
\operatorname{D}f=\left(
    \begin{array}{cccccc}
         0 & 0 & 0 & 1 & 0 & 0 \\
         0 & 0 & 0 & 0 & 1 & 0 \\
         0 & 0 & 0 & 0 & 0 & 1 \\
         A_{x,x} & A_{x,y} & A_{x,z} & 0 & 0 & 0 \\
         A_{y,x} & A_{y,y} & A_{y,z} & 0 & 0 & 0 \\
         A_{z,x} & A_{z,y} & A_{z,z} & 0 & 0 & 0 \\
    \end{array}
\right).
\end{equation}
Here, the subindices $x,y,z$ denote the spatial components of the position
of Halley's comet, with respect to which the partial derivatives are computed.
For $A_{x,x}$ we obtain
\begin{equation}
A_{x,x} = \sum_{i=1}^N \left[
    \frac{3 Gm_i (x_H-x_i)^2}{r_{H,i}^5}-\frac{G m_i}{r_{H,i}^3}\right],
\end{equation}
and have analogous expressions for $A_{y,y}$ and $A_{z,z}$, which are
obtained by replacing $x$ by $y$ and $z$, respectively. Likewise, $A_{x,z}$ is
given by
\begin{equation}
A_{x,z} = A_{z,x} = \sum_{i=1\neq i}^N \left[
    \frac{3 Gm_i (x_H-x_i) (z_H-z_i)}{r_{H,i}^5}\right],
\end{equation}
and similar expressions can be obtained for $A_{x,y}$ and $A_{y,z}$ by replacing
appropriately the spatial variables. The initial conditions for the variational
equations correspond to the identity matrix $\xi_0 = \mathbb{1}_6$.

During the numerical integration of the equations of motion and the
variational equations, at fixed time intervals $t_k = k \cdot \Delta t$
($k=1,2,\ldots$) we perform a QR-factorization of the solution of the variational
equations~\citep{pikovsky2016lyapunov}, i.e., we write $\xi(t_k)=Q\cdot R$. Here, $Q$ is an orthogonal
matrix and $R$ is an upper triangular matrix with positive diagonal
elements. The diagonal elements of $R$ allow to
compute all independent growth factors, from which the $l$-th Lyapunov exponent
is computed at time $t_k$ as
\begin{equation}
\label{Lyaps}
\lambda_l(t_k)=\sum_{m=1}^k\log\big(R_{ll}(t_m)\big)/t_k.
\end{equation}
The components of the $Q$ matrix are then substituted into $\xi(t_k)$, which are the
new initial conditions of the variational equations for the next integration.
In our integrations, we use $\Delta t = T_\mathrm{Me}/2$, where $T_\mathrm{Me}$ is
Mercury's orbital period; this choice allows us to perform a Fourier analysis
and resolve frequencies as large as Mercury's orbital frequency.

The equations of motion (\ref{Eq:Hvel}) and~(\ref{Eq:Hacc}) can be derived from a
Hamiltonian. Hence, the symplectic structure imposed by the Hamiltonian implies
that the Lyapunov spectrum satisfies
\begin{equation}
	\sum_{l=1}^6\lambda_l = \operatorname{Trace}(\operatorname{D}f) = 0.
	\label{eq:symplectic}
\end{equation}
This property of the Lyapunov spectrum will also be used below as a test of our
numerical results.

\section{Results}
\label{Sec4}

\subsection{Osculating orbital elements}
\label{sec4:orb_ele}

\begin{figure}
	\includegraphics[width=\columnwidth]{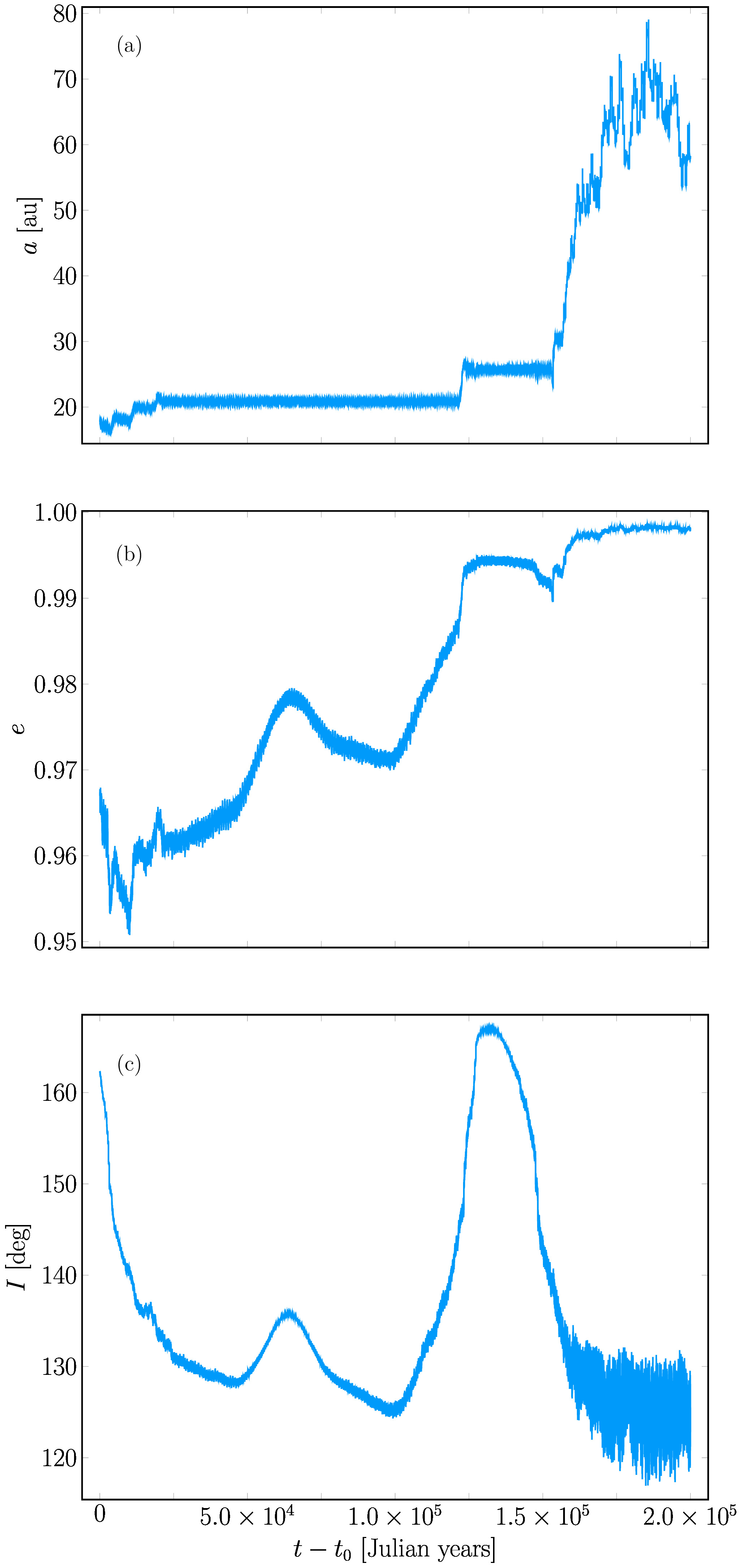}\\
	\caption{ (a)~Semi-major axis, (b)~eccentricity and
    (c)~inclination of Halley's comet orbit as a function of time,
    obtained using the nominal initial conditions defined at $t_0 = 2449400.5$~JD
    (February 17, 1994).
    \label{fig:fig1}
    }
\end{figure}

We begin our description of the dynamics of Halley's comet considering the
variation in time of the osculating orbital elements.
For the initial conditions of all Solar System bodies we use the data from
the JPL's Horizons on-line ephemeris service~\citep{giorgini1996jpl}
considering February 17th, 1994, 00:00:00.0 (TDB). This
date ($t_0 = 2449400.5$~JD) is the reference epoch for JPL's current solution
for Halley's orbit~\citep{giorgini1996jpl}. All initial conditions
are referred to the Solar System's barycenter; the $x$--$y$ plane corresponds
to the mean ecliptic at the J2000.0 epoch. The initial conditions for
all planets correspond to the respective planetary barycenter considering their
orbiting moons, except for Mercury and Venus; similarly to DE430/431
ephemerides~\citep{folkner2014planetary}, in our model the orbits of the Earth
and the Moon are integrated as separate bodies. The $G m_i$
values of all bodies are taken from the JPL DE430/431 ephemerides documentation
as well; see Table~\ref{tab:tab1}. Using these initial conditions, in
Figs.~\ref{fig:fig1} we plot the results obtained for
Halley's semi-major axis $a$, eccentricity $e$ and inclination $I$
as a function of time; the zero in the time axis represents the epoch $t_0$
of the initial conditions.

\begin{table}
	\centering
    \caption{$Gm_i$ values, current values of the orbital periods $T$ (yr)
	and frequencies $f=1/T$ (yr$^{-1}$) of the planets and 1P/Halley comet,
	used in our calculations. The planetary orbital periods
    were adapted from table 8.7 in \citet{urban2014explanatory};
    Halley's orbital period was retrieved from HORIZONS~\citep{giorgini1996jpl}.
	For the Sun and the Moon, the corresponding $ Gm_i$ values are
	$G m_0 = 132712440041.93938$~km${}^3$s${}^{-2}$ and
    $ G m_4 = 4902.800066$~km${}^3$s${}^{-2}$ \citep{giorgini1996jpl}.
    }
	\label{tab:tab1}
	\begin{tabular}{lccc}
		\hline
		Body & $G m_i$              & Orbital period & Orbital frequency\\
		     & (km${}^3$s${}^{-2}$) & (yr)           & (yr$^{-1}$) \\
		\hline
		Mercury & 22031.78      & 0.241   & 4.152 \\
		Venus   & 324858.592    & 0.615   & 1.625 \\
		Earth   & 398600.435436 & 1.000   & 1.000 \\
		Mars    & 42828.375214  & 1.881   & 0.532 \\
		Jupiter & 126686534.911 & 11.868  & 0.084 \\
		Saturn  & 37931207.8    & 29.437  & 0.034 \\
		Uranus  & 5793951.322   & 84.048  & 0.012 \\
		Neptune & 6835099.5     & 164.891 & 0.006 \\
		Halley  & NA            & 75.316  & 0.013 \\
		\hline\\[1ex]
	\end{tabular}
\end{table}

Within the first few thousand years, Figs.~\ref{fig:fig1} display some sudden changes
in the semi-major axis, eccentricity and inclination of Halley's orbit. These events display
the accumulation effects of
close approaches with Jupiter; in particular, an important one occurs at
$t-t_0 \simeq 2925$~yr, in which Halley's comet enters the Hill radius of Jupiter
reaching a minimum distance of $\sim 0.445$~$\rho_\textrm{J}$, where
$\rho_\textrm{J} \approx 0.355$~au is Jupiter's Hill radius. This event was already pointed out
by~\citet{munoz2015chaotic}. \citet{boekholt2016origin} noticed an important role
played by Venus during the first $\sim 3\times10^3$~yr; we confirm the occurrence
of some approaches to Venus during that time, the closest one occurring around
$t - t_0\simeq 732$~yr, with the comet being within a distance
$\sim 8.08 \rho_\textrm{V}$ from the planet, where
$\rho_\textrm{V} \approx 6.76\times10^{-3}$~au is Venus' Hill radius.
Those close approaches lead to sudden increments of the semi-major axis, which
display oscillations of $\sim 1$~au. Eventually, Halley's comet semi-major axis
oscillates in a seemingly steady form around $a\approx 21$~au for a long time
($\sim10^5$~yr), avoiding close encounters with Jupiter. During that time, the
eccentricity and inclination still display changes, in particular a marked
bump reaching $e\sim 0.98$ and $I\sim 135^\circ$,
which we interpret as the effect of the accumulation of secular
instabilities. This bump is followed by an abrupt increase in the eccentricity
and inclination, this time due to a close approach with the Sun.
Beyond $t - t_0 \gtrsim 1.1\times 10^5$~yr, these instabilities allow new close
encounters with Jupiter and the Sun which make the eccentricity attain rather large
values, $e \sim 0.998$, induce a clear increment of the semi-major axis,
and a drastic reduction of the inclination, which also displays larger
fluctuations. The large values of the eccentricity attained
suggest the possibility of an eventual ejection
of Halley's comet from the Solar System; we shall return later to this observation.

\begin{figure}
	\includegraphics[width=\columnwidth]{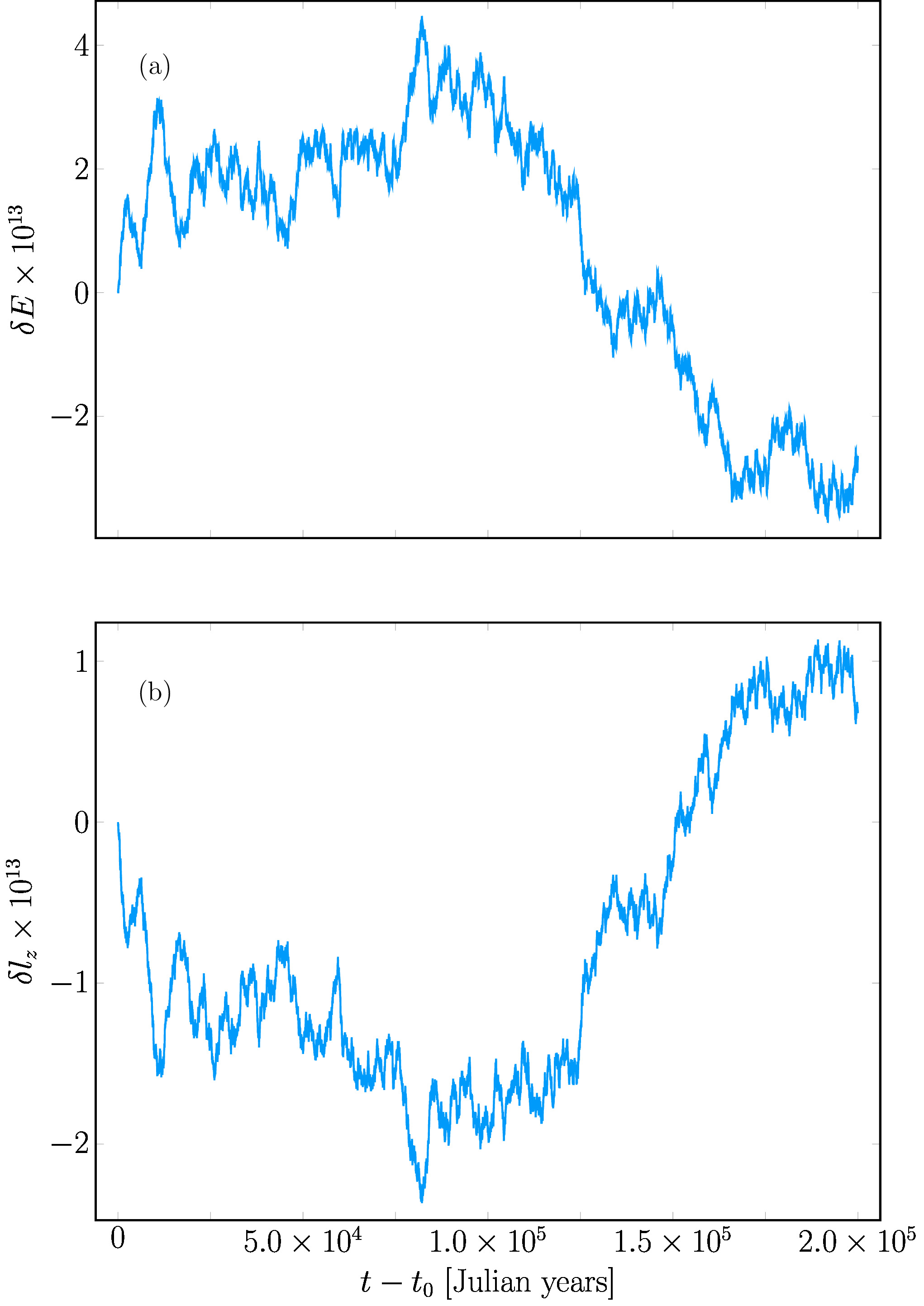}
	\caption{(a) Relative variation of the total energy and
    (b)~$z$-component of the total angular momentum of the Solar System
    (Sun and planets) as a function of time.}
    \label{fig:fig2}
\end{figure}

Regarding the quality of our integration, in Figs.~\ref{fig:fig2} we display
the relative change of the total energy of the Solar System
with respect to its initial value, $\delta E$, and the corresponding
relative change of the $z$-component of the angular
momentum of the Solar System, $\delta l_z$, as functions of time.
As illustrated, we obtain $|\delta E| \leq 3\times 10^{-13}$ and
$|\delta l_z| \leq 1.1\times 10^{-13}$. The overall behavior of these relative
changes resembles a random walk; this is a consequence of the error per time
step of our integrations being essentially the machine round-off error.
These figures show that accumulation of those errors over the
integration time are small and close to the initial energy
and angular momentum values, which are the dynamical constraints
imposed by the integrals of motion. While these results are independent
of the presence of Halley's comet, which is treated as a massless particle, they
do influence the motion of the comet through their presence in Eq.~\ref{Eq:Hacc};
small errors in the preservation of the integrals of motion in the $N$-body
problem are considered more trustworthy numerical results \citep{zwart2014minimal}.

\subsection{Fourier analysis: Power spectrum}
\label{sec:fourier}

\begin{figure}
	\includegraphics[width=\columnwidth]{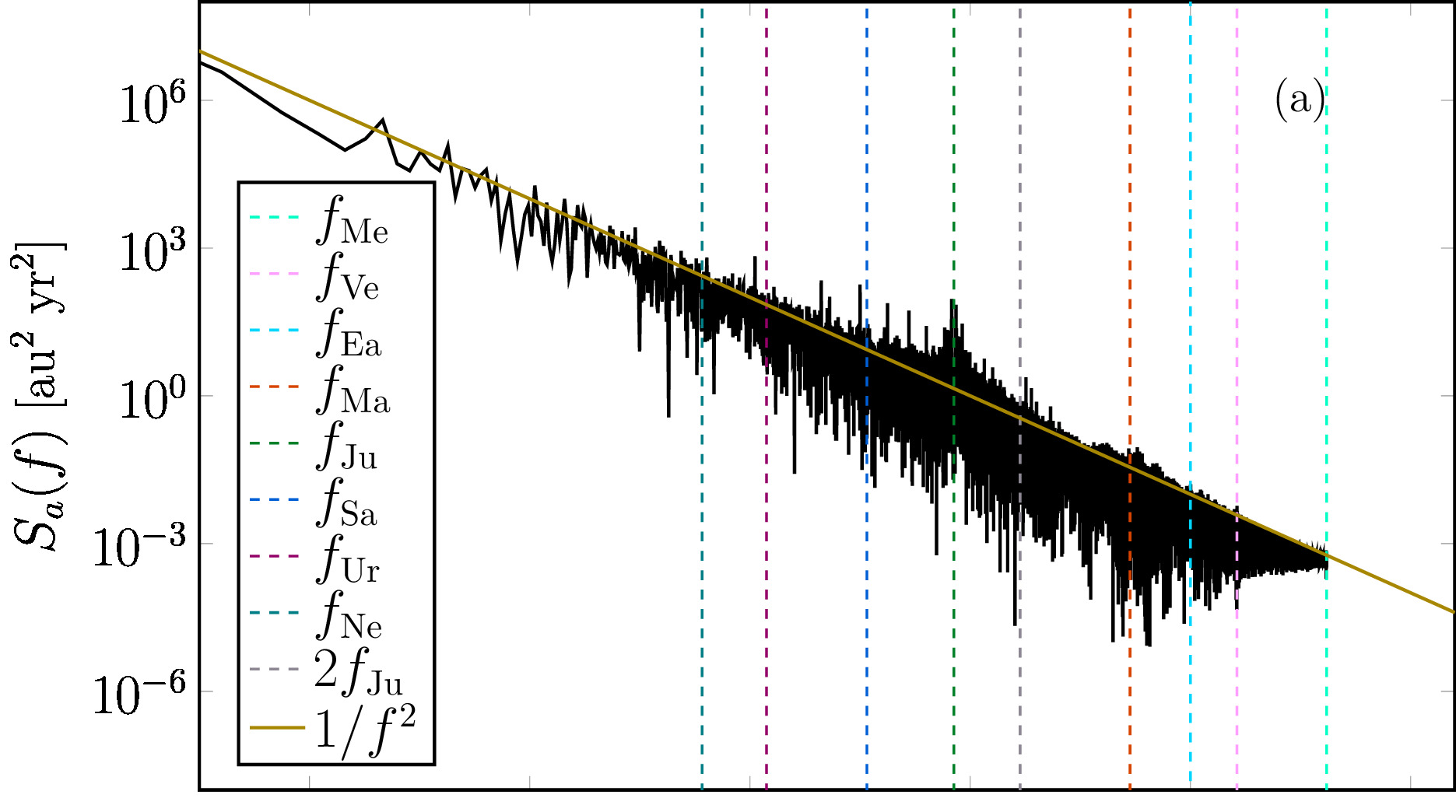}\\
    \includegraphics[width=\columnwidth]{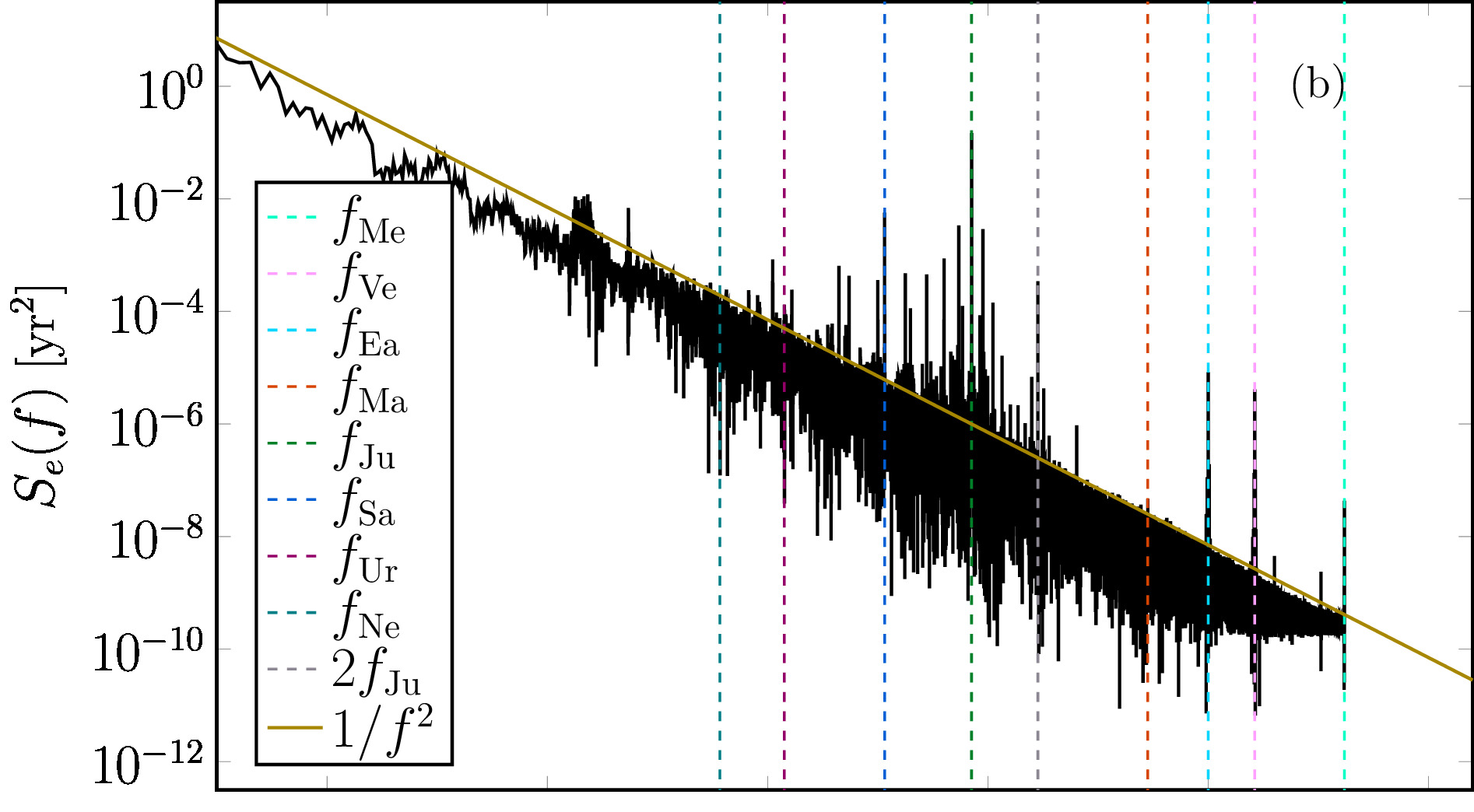}\\
    \includegraphics[width=\columnwidth]{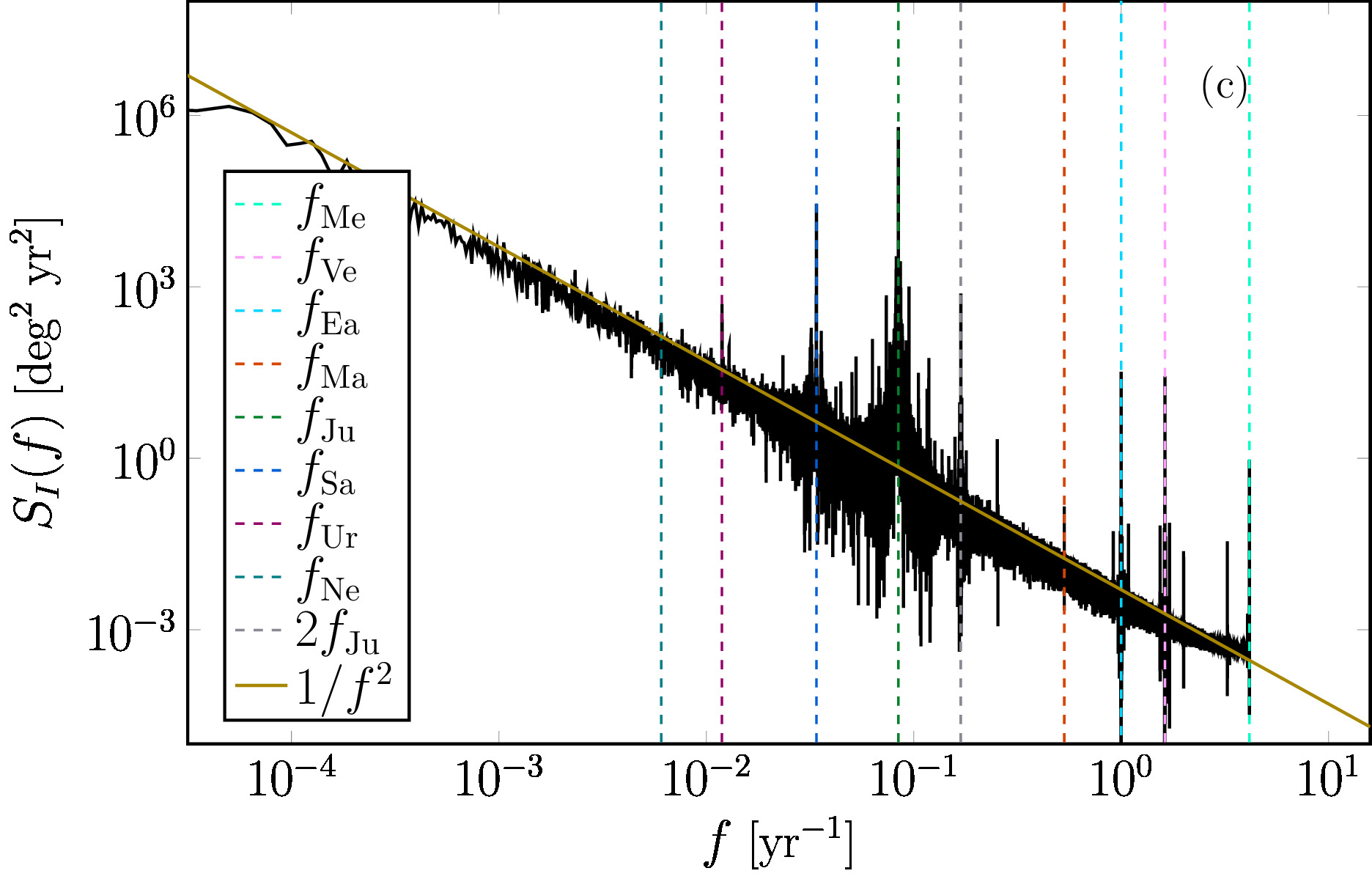}
    \caption{Power spectrum (a)~$S_a(f)$, (b)~$S_e(f)$ and (c)~$S_I(f)$ (black)
    associated to Halley's osculating semi-major axis, eccentricity and inclination,
    computed over $2\times10^5$~yr. Vertical dashed
    lines correspond to current planetary orbital frequencies as indicated in
    the legend. The figures include the $1/f^2$ decay to separate any interesting
    feature from the bulk.}
    \label{fig:figs3}
\end{figure}

In order to further understand Halley's dynamics in more detail, we have computed
the discrete Fourier Transform associated to the time series defined
by the Halley's osculating orbital elements displayed in Figs.~\ref{fig:fig1},
from which we obtain the associated power spectra.
We denote by $S_a(f)$, $S_e(f)$ and $S_I(f)$ the power spectra associated to
Halley's semi-major axis, eccentricity and inclination, respectively.
In Figs.~\ref{fig:figs3} we present the results as a function of the
frequency, in log-log scale. In these figures, the vertical dashed lines correspond
to the current planetary orbital frequencies, including also $2f_\mathrm{Ju}$;
table~\ref{tab:tab1} includes the
values of the planetary orbital periods and the corresponding orbital
frequencies. In these figures we have additionally indicated the $1/f^2$ decay,
which serves to distinguish any important feature from the bulk. The figures
clearly show a broadband component which are a clear manifestation
of the chaotic dynamics~\citep{sussman1988numerical} of comet Halley.

Figure \ref{fig:figs3}(a) shows an important accumulation of peaks
located around Jupiter's orbital frequency $f_\mathrm{Ju}$, which clearly form a
bump. There is also a much smaller bump around Mars frequency $f_\mathrm{Ma}$, and
a small peak coinciding with the present orbital frequency of Venus,
$f_\mathrm{Ve}$. These features reflect the increment of the
semi-major axis displayed by Halley's orbit, illustrated in Fig.~\ref{fig:fig1}(a),
due to different close approaches.

Likewise, in Figs.~\ref{fig:figs3}(b) and \ref{fig:figs3}(c),
$S_e(f)$ and $S_I(f)$ display some peaks that match the planetary
orbital frequencies of all planets, except for Uranus and Neptune in the
case of $S_e(f)$. The strongest peaks with respect to the background
(the $1/f^2$ line)
correspond to the frequencies clustered around Jupiter's current frequency,
though the frequencies of Saturn ($f_\mathrm{Sa}$), Venus ($f_\mathrm{Ve}$)
and the Earth ($f_\mathrm{Ea}$) seem also to play a significant role.
We also note the approximate $1:6$ commensurability among the frequencies of
Jupiter and Halley's comet, i.e., $f_\mathrm{Ju}\approx 6f_\mathrm{Ha}$
\citep{sekhar2014resonant}.
The appearance of some peaks between the frequencies of Mars and Jupiter is
associated to harmonics of this approximate resonance; notice in particular
that $S_I(f)$ displays contributions of the $2:12$ and $3:18$
harmonics. Using the relative height of the resonance peaks
with respect to the local power spectrum signal, Jupiter, Saturn, Venus
and the Earth seem to be the main perturbers of Halley's eccentricity and
inclination.

\subsection{Lyapunov spectrum}
\label{sec:lyapunov}

\begin{figure}
	\includegraphics[width=\columnwidth]{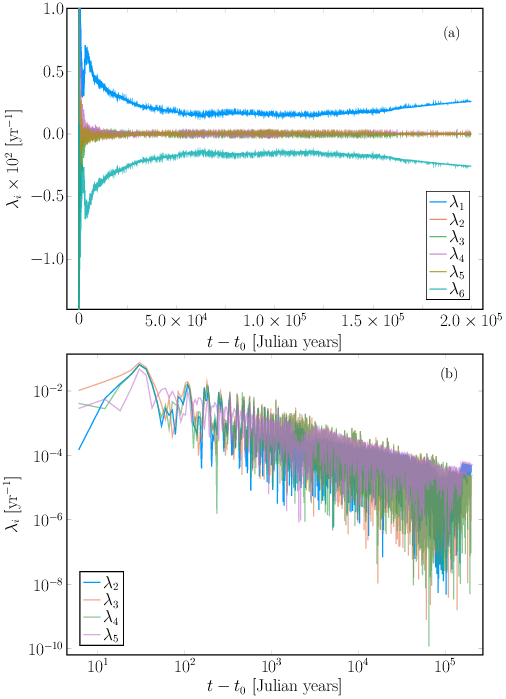}
    \caption{Time dependence of the Lyapunov spectrum: (a) All exponents are
    displayed in linear scale; (b) absolute value of all
    exponents except for the maximum Lyapunov exponent $\lambda_1$, and its
    time-reversal partner, $\lambda_6$, in log-log scale. We notice that the exponents included
    in (b) display a convergence to zero in time, seemingly as $t^{-1/2}$.}
    \label{fig:fig4}
\end{figure}

We consider now the Lyapunov spectrum of Halley's orbit.
In Fig.~\ref{fig:fig4}(a) we present the time dependence of each of the six
Lyapunov exponents of Halley's orbit using a
linear scale; Fig.~\ref{fig:fig4}(b) displays in a log-log scale the absolute value
of the Lyapunov exponents, excluding the maximum one, $\lambda_1$, and the minimum
one, $\lambda_6$. Clearly, only $\lambda_1$ and its time-reversal partner
$\lambda_6$ seem to attain asymptotically non-zero values.
The upper bound of the absolute value of the remaining exponents tends
to zero as a power law in time, seemingly as $t^{-1/2}$, though important
fluctuations are observed. That is, along the orbit of the comet there is one
(time-dependent) direction in phase space where the instabilities and thus chaos
are particularly strong, another one corresponding to its time-reversal (stable)
counterpart, and the remaining independent directions behave asymptotically
as neutral directions
with respect to long-term stability. Therefore, small deviations in the initial
conditions which are not perfectly perpendicular to the single unstable direction
will display sensitivity to the initial conditions.

\begin{figure}
	\includegraphics[width=\columnwidth]{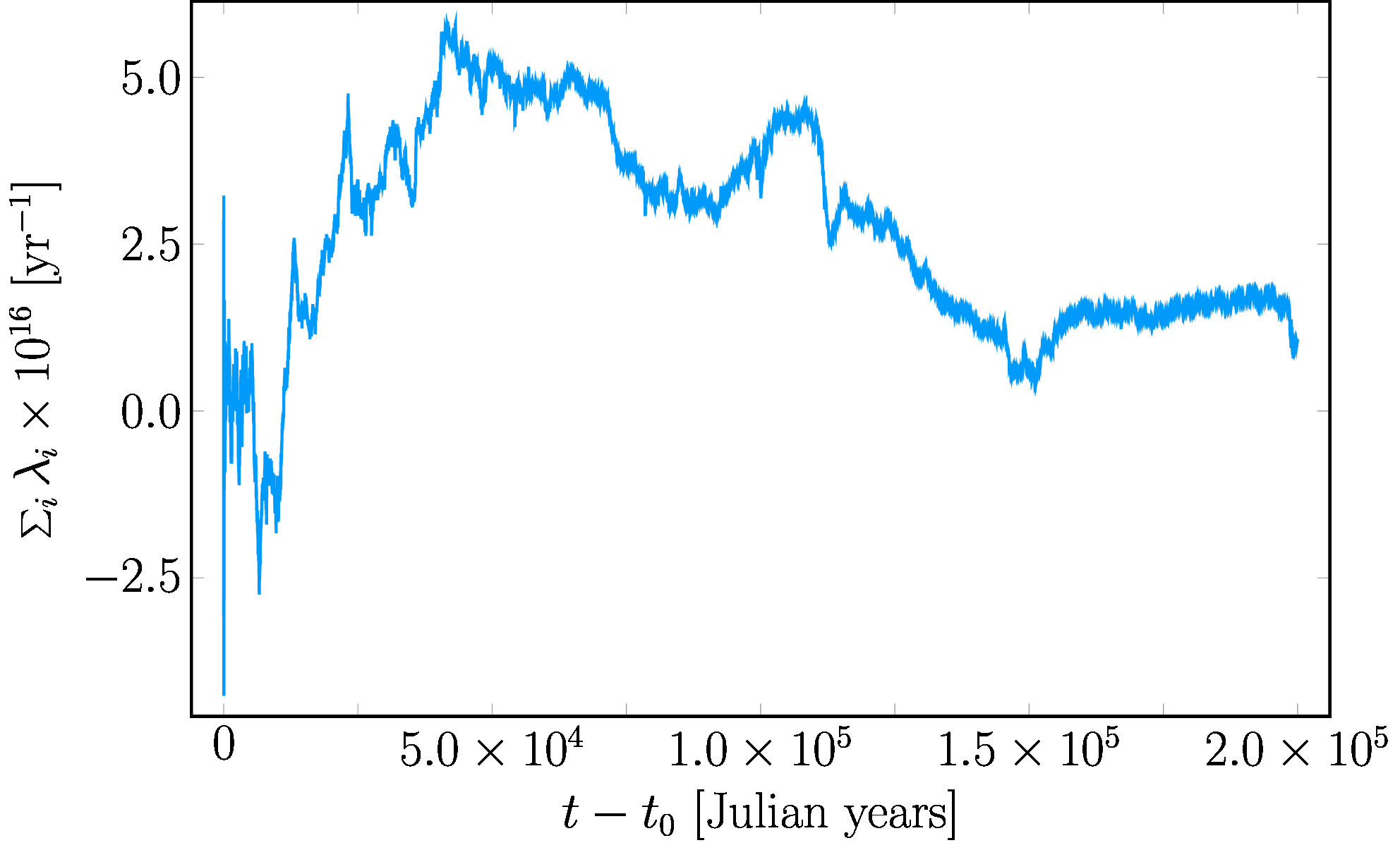}
    \caption{Time dependence of the sum of the Lyapunov exponents, as a test
    of the preservation of the symplectic structure of the equations of
    motion by our integration method.}
    \label{fig:fig5}
\end{figure}

We address now the validity of the calculations of the Lyapunov exponents
presented. In Fig.~\ref{fig:fig5} we display the time dependence of the sum of
the Lyapunov exponents. As noted above, Eq.~(\ref{eq:symplectic}) permits to
check how the symplectic structure of the equations of motion for Halley's comet
is preserved by the integration.
Figure~\ref{fig:fig5} illustrates that the variations of $\sum_i \lambda_i$ remain
very close to zero for all times, and are essentially due to the accumulation
of round off errors in each time-step of the integration.
We emphasize that Taylor's method is not symplectic;
yet, by using polynomials of high-enough order and a small
enough tolerance $\varepsilon$, the method preserves numerically the symplectic
structure.

Figure~\ref{fig:fig4}(a) shows that, despite of the fact that the maximal
Lyapunov exponent $\lambda_1$ is non-zero, it has not yet attained
convergence during the integration time considered. This is apparent by the
increasing tendency displayed during the last $4\times 10^4$~yr of our integration.
In order to have an estimate of the Lyapunov time, we consider
the average value of the maximum Lyapunov exponent for
times $t - t_0 \ge 5\times 10^4$~yr. This criterion
is chosen because the changes of $\lambda_1$ seem to remain relatively constant
during this time interval, except for the last $4\times 10^4$~yr
of the orbit. We obtain $\overline{\lambda_1} = 1.777\times 10^{-3}$~yr$^{-1}$,
with the minimum and maximum values corresponding to $1.424\times10^{-3}$~yr$^{-1}$
and $2.597\times 10^{-3}$~yr$^{-1}$. From the average value, our estimate of the
Lyapunov time for Halley's comet is $\tau_L=1/\overline{\lambda_1} \simeq 562$~yr.
This result is larger than the values reported by \citet{munoz2015chaotic},
$\tau_L \sim 70$~yr, and \citet{boekholt2016origin}, $\tau_L \sim 300$~yr.

\section{Discussion}
\label{Sec5}

In view of the different results obtained for $\lambda_1$,
it is worth describing the differences of all models that are under comparison. First,
the initial conditions used are all different as well as the numerical methods employed.
These are important due to the underlying chaotic dynamics of comet Halley.
The Solar System models used are also different, though all include only
Newtonian gravitational interactions: \citet{munoz2015chaotic}
include all planets except for Mercury, whose mass is added to the Sun, five
dwarf planets and five minor bodies. In turn, \citet{boekholt2016origin} use all
planets as we do, except for the inclusion of the Moon. The time scales
spanned by the calculations are also different, ours being the longest.
Finally, the actual methods used to compute the maximal Lyapunov
exponents are different. As noted by \citet{boekholt2016origin},
\citet{munoz2015chaotic} use the iterative scheme proposed
by \citet{benettin1976kolmogorov}, while they measure directly the rate of
exponential growth from the separation of the two orbits. We note that
both groups use fiducial orbits in their calculations. Our method is based on
the numerical integration of the first variational
equations~\citep{benettin1980lyapunovII} which avoids the use of any fiducial
orbit and thus is not subject to artificial overestimations~\citep{2001AJ....121.1171T},
though this method is computationally more demanding.

The lack of convergence of $\lambda_1$ (and $\lambda_6$), manifested in the last
$4\times 10^4$~yr of our integration, seems reminiscent of early findings
for the maximum Lyapunov exponent of the Jupiter family
comets~\citep{tancredi1995dynamical}. There, it was observed that some
Lyapunov exponents seemed to converge up to a certain time, followed
by a sudden change associated to the orbit crossing a partial dynamical
barrier (cantorus) in phase space, which increased the value of the calculated
Lyapunov exponent. This behavior was attributed to (recurrent)
close encounters of the comets with Jupiter~\citep{tancredi1995dynamical}
which allowed the comets to escape from a phase-space region with reduced
diffusion (sticky region) to one with enhanced
diffusion~\citep{contopoulos1989lyapunov}. We do observe the impulsive close
approaches with different bodies in our simulations, but still with no convergence
of the Lyapunov exponent.

Convergence in the limit of infinite time is expected by Osedelet's
theorem~\citep{Oseledet1968multiplicative}.
Disregarding the lack of its applicability ---since there are
clear indications of irregular motion---, the lack of
convergence of the Lyapunov exponents could be attributed to the finite,
and perhaps short, integration time considered. While this is certainly a limitation,
the Lyapunov time obtained is much shorter than the total time-scale of our
integration, from which convergence could be expected. Another alternative
to understand the lack of convergence is
that comet Halley will eventually escape. This is possible through the
accumulation of close encounters which cause sudden changes in the
osculating orbital elements and in the value of the Lyapunov exponents,
as shown in Figs.~\ref{fig:fig1} and~\ref{fig:fig4}. Noticing that
towards the end of our integration the computed orbit displays values for
the eccentricity that are very close to 1, close approaches with the Sun
are expected essentially once per orbital period. These encounters are an
efficient mechanism for ejection,
which leads us to conclude that comet Halley will eventually escape from the
Solar System. From our integration, the closest approach of comet Halley
to the Sun occurs at $t-t_0\approx 1.776\times10^5$~yr, at a distance
$\sim 0.134$~au, that is, about $1/3$ the current semi-major axis of Mercury,
i.e. $\sim 30~R{}_\odot$.

\begin{figure}
    \includegraphics[width=\columnwidth]{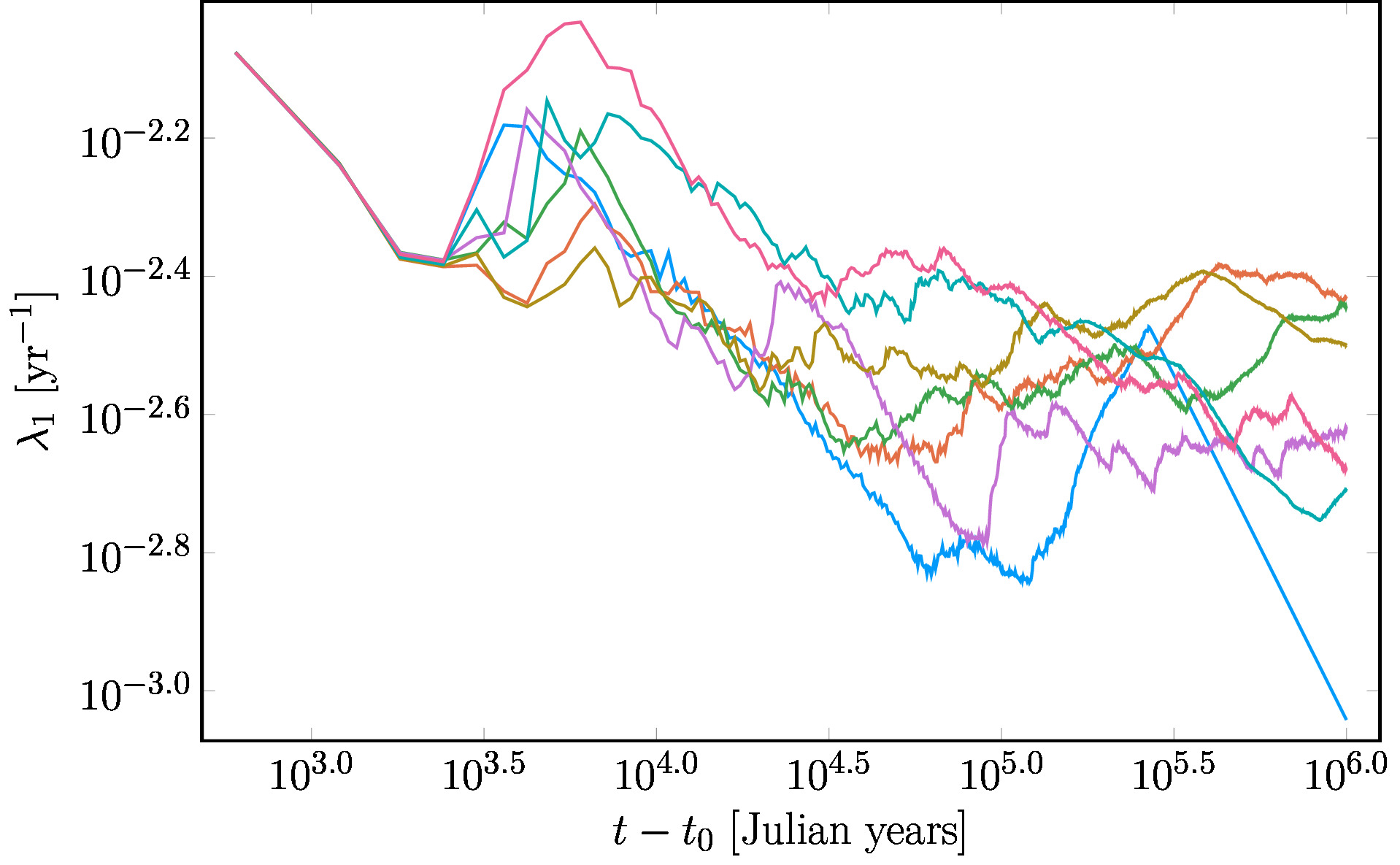}
    \caption{Time dependence of maximum Lyapunov exponent $\lambda_1$, during
    $1\times 10^6$~yr, of the nominal Halley orbit (blue line), and
    6 nearby initial conditions.}
    \label{fig:fig6}
\end{figure}

If the comet escapes, the dynamics become integrable in the asymptotic region,
as the comet will only feel the interaction with the Sun, with a mass corrected to include
the other Solar System bodies, i.e., after escape the dynamics corresponds essentially
to the two-body problem.
In this case, the complex dynamics displayed by
Halley's comet are only temporary, limited in time, and in that sense a
manifestation of chaotic scattering or transient
chaos~\citep{SpecialIssueChaos1993,LaiTelBook2011}. Then, all Lyapunov
exponents converge to zero due to integrability of the asymptotic dynamics,
where the particle escapes as $t^{2/3}$. We note that this is in agreement with
Osedelet's theorem. Accordingly, finite-time estimations of the Lyapunov time $\tau_L$
in the vicinity of comet Halley's nominal orbit display a lack of convergence, different time
dependence and values, being highly dependent on the initial conditions as well
as the propagation of numerical errors. Since convergence is not reached,
the computed values of $\tau_L$ are transient and hard to interpret,
and it is not clear to us if they entail any significance.

\begin{figure}
    \includegraphics[width=\columnwidth]{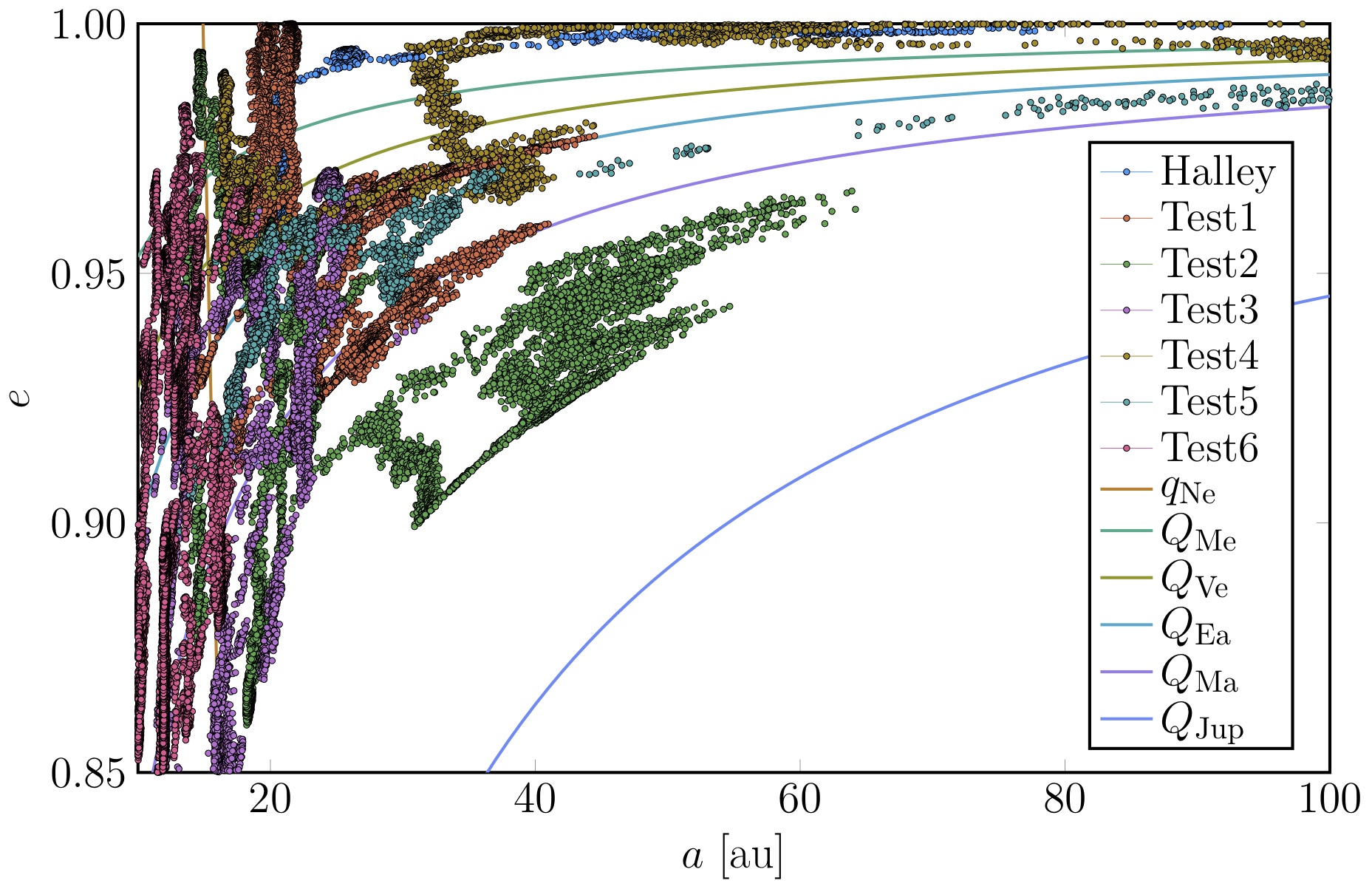}
    \caption{Evolution of Halley's comet semimajor axis and eccentricity $(a, e)$
        at each perihelion passage, computed over $1\times10^6$~yr, for
        each of the orbits in Fig. \ref{fig:fig6}. Solid curves
        show the values of the pair $(a, e)$ corresponding, respectively, to the
        perihelia ($Q$) and aphelia ($q$) of Mercury, Venus, Earth, Mars,
        Jupiter and Neptune. }
    \label{fig:fig7}
\end{figure}

\begin{figure}
    \includegraphics[width=\columnwidth]{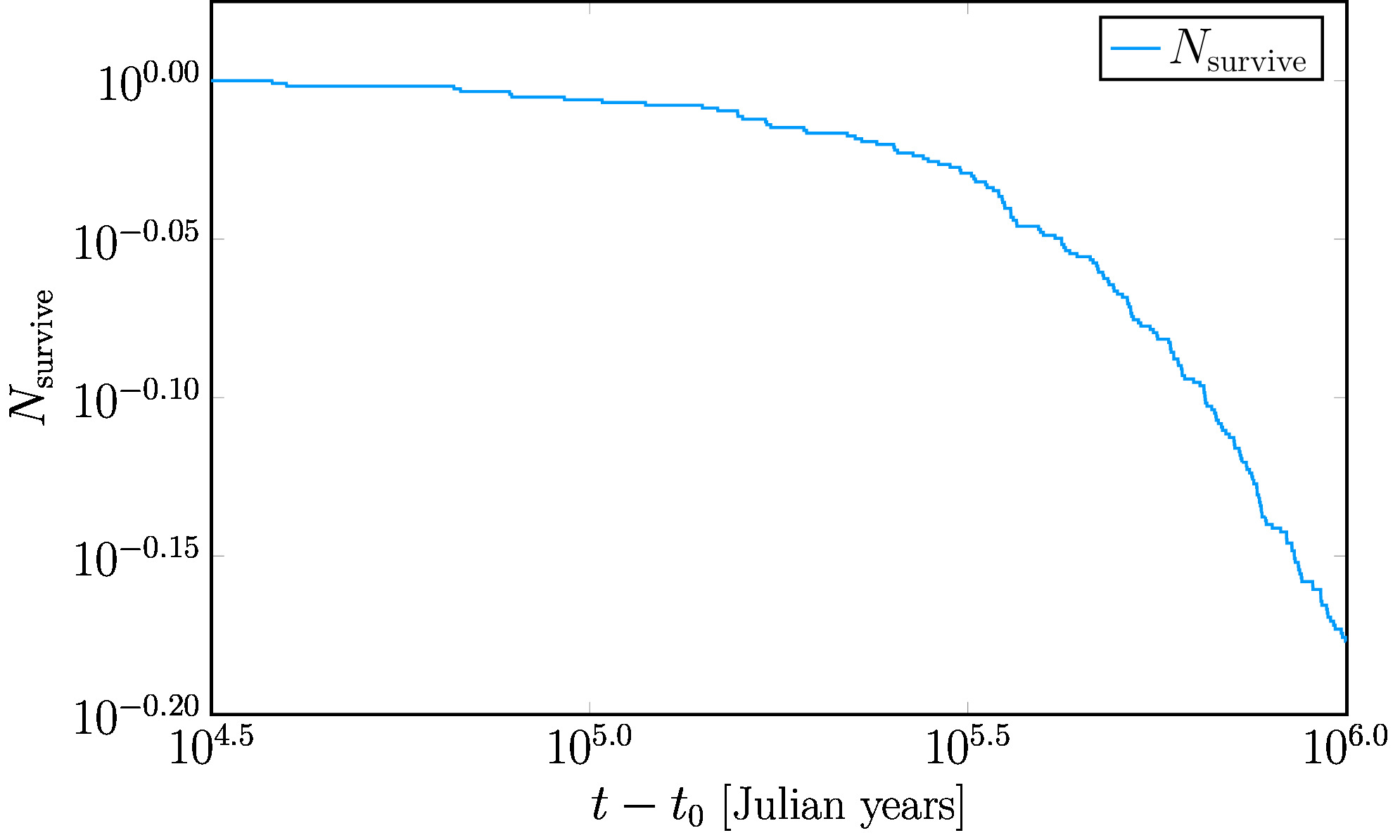}
    \caption{Decay of the survival probability of 508 initial
    conditions considered at random, within a sphere of $150$~km centered
    at the nominal position of Halley's comet.}
    \label{fig:fig8}
\end{figure}

To illustrate this conclusion, we integrate comet Halley's nominal orbit, as well as
6 nearby initial conditions, during $1\times 10^6$~yr. Figure \ref{fig:fig6}
shows the maximum Lyapunov exponent $\lambda_1$ associated to each of those
initial conditions. The perturbed initial conditions differ by $150$~km in the
initial position with respect to the nominal orbit. We observe that
the nominal orbit (light blue) does indeed escape after $\sim 2\times 10^5$~yr,
when the comet's orbit becomes hyperbolic. This is manifested in Fig. \ref{fig:fig6}
as $\lambda_1$ converging to zero, seemingly as $t^{-1}$. From our results, the
escaping time for the nominal initial conditions is about $2.7\times10^5$~yr.
Furthermore, we note that while the rest of the integrated orbits
shown in Fig. \ref{fig:fig6} do not escape the Solar System within
$1$~Myr, their maximum Lyapunov exponents do not display convergence either.
We also notice in Fig. \ref{fig:fig6} that the maximum Lyapunov exponent
of the computed trajectories have the same value up to $\sim 2.72\times 10^3$~yr,
and afterwards they evolve independently. This is related to a close approach
to Jupiter, in which the comet almost enters the Hill sphere of the planet.
In Fig.~\ref{fig:fig7} we plot, in the $a$--$e$ plane, the evolution of
the osculating elements at perihelion passage for each of the orbits shown in
Fig. \ref{fig:fig6}. Clearly, there is a long-term sensitivity of the osculating
elements to perturbations of the initial condition, as evidenced by the different
regions visited, where we observe that three of the computed trajectories
reach $100$~au for the semi-major axis with quite high eccentricity.

In order to characterize the rate of escape of the vicinity of comet Halley's
orbit, we consider 508 initial conditions close to the nominal orbit,
and integrate them up to $1\times 10^6$~yr. Figure~\ref{fig:fig8} shows the
fraction of initial conditions $N_\mathrm{survive}$ which have not escaped
at time $t$. The sampled initial conditions are obtained by varying
the initial position of the test particle within a radius of
$150$~km, which is essentially the uncertainty of comet Halley's semimajor axis
\citep{1986A&A...163..246L}. The escape condition requires that the comet's
orbit is hyperbolic in a Keplerian sense, that is, that $e>1$ and its Keplerian
energy is positive. We notice that these results are consistent with the
structure of the survival maps
of Halley's comet~\citep{munoz2015chaotic}, and manifest that the escape time is
highly sensitive to the initial conditions and the propagation of numerical
errors~\citep{1998BST,BBMSPRE71p036225}.
We thus conclude
that comet Halley will likely escape from the Solar System, and $\lambda_1$
will display non-converging transient behavior until the comet escapes,
after which $\lambda_1\to 0$.

\section{Conclusions}
\label{Sec6}

In this paper, using a purely Newtonian model for the Solar System, we have
computed numerically the orbit of comet 1P/Halley for $2\times10^5$~yr, using for
the initial conditions those provided by JPL's Horizons system, corresponding
to February 17th, 1994, 00:00:00.0 (TDB). The orbit computed displays
important abrupt changes due mainly to close approaches with Jupiter, though
other planets influence importantly the comet's trajectory. Secular resonances
and close approaches with the Sun are also manifested.

We have studied the power spectra associated to Halley's
semi-major axis $S_a(f)$, eccentricity $S_e(f)$ and inclination $S_I(f)$. The
power spectra display a broadband component, which is consistent
with the chaotic dynamics~\citep{sussman1988numerical} of Halley's orbit. The
results for $S_e(f)$ and $S_I(f)$ manifest peaks associated to the
current planetary frequencies, which are dominated in particular by Jupiter,
Venus, Saturn and the Earth, according to the relative strength of the
associated peaks with respect to the $1/f^2$ decay of the bulk. In addition,
the $1:6$ resonance with Jupiter seems to be
relevant~\citep{sekhar2014resonant} as well as some
of its harmonics. Therefore, these results support that
close approaches with Jupiter influence importantly
Halley's orbital elements, but also resonant interactions with other planets,
including Venus, as it has been recently suggested~\citep{boekholt2016origin}.

We have also presented results on the Lyapunov spectrum of the comet, which has
been computed integrating the variational equations.
For the nominal orbit of
comet Halley, we obtained for the average maximum Lyapunov exponent
$\overline{\lambda_1} \approx 1.777\times 10^{-3}$~yr$^{-1}$, from which we
estimate the Lyapunov time $\tau_L = 1/\overline{\lambda_1} = 562$~yr.
Notice that for this value of $\tau_L$, an
initial deviation $\delta_0=150$~km requires $\sim 2100$~yr to grow to
the radius of the Earth. This value is larger than the previously reported
\citep{munoz2015chaotic,boekholt2016origin},
and it roughly agrees with the interval of confirmed observations of the comet.
Yet, we point out that our results, up to
$2\times10^5$~yr, do not display convergence for $\lambda_1$,
which is also manifested in integrations of initial conditions
close to the nominal orbit of the comet. We interpret
the lack of convergence of the maximum Lyapunov
exponent as a signature of transient chaos, which will lead to an eventual
ejection of the comet from the Solar System. Numerical computations
spanning $1$~Myr confirm the escape of the nominal orbit comet as well as other
neighboring trajectories.

Aside from the non-zero values of $\lambda_1$ and $\lambda_6$,
the latter being the time-reversal associate
of $\lambda_1$, and the fact that these exponents have not converged
by the end of our simulations,
the rest of the Lyapunov exponents seem to converge to zero
asymptotically as $t^{-1/2}$. We have
presented results on the total energy of the Solar System, the $z$-component
of its total angular momentum, and the sum of all Lyapunov exponents, which
provide good confidence checks on the numerical results obtained.

In our work, as well as in previous
works~\citep{munoz2015chaotic,boekholt2016origin}
no relativistic corrections in Mercury's orbit have been taken into account,
nor during Halley's passages through perihelia. Similarly, the non-gravitational
cometary accelerations~\citep{Marsden1973} have been disregarded. It is clear
that those contributions may influence the possible ejection of Halley's
comet from the Solar System; these questions remain open.

\section*{Acknowledgments}
We are thankful to B\'arbara Pichardo and Marco Mu{\~n}oz-Guti{\'e}rrez for
discussions and comments.
We acknowledge financial support provided by UNAM-PAPIIT IG-100616 and
computing resources provided by LANCAD-UNAM-DGTIC-284.
J.A. P\'erez-Hern\'andez acknowledges financial support provided by CONACyT.




\bibliographystyle{mnras}
\bibliography{halley} 








\bsp	
\label{lastpage}
\end{document}